\documentclass[twocolumn, superscriptaddress, floatfix]{revtex4-2}

\usepackage[utf8]{inputenc}
\DeclareUnicodeCharacter{00A0}{ }
\usepackage{amsmath}
\usepackage{amssymb}
\usepackage{amsthm}
\usepackage{bbold}
\usepackage{mathtools}
\usepackage{physics}
\usepackage[makeroom]{cancel}
\usepackage{stmaryrd}
\usepackage{graphicx}
\usepackage{color}
\usepackage[english]{babel}
\usepackage{stmaryrd}
\usepackage{microtype}
\usepackage{float}
\usepackage{siunitx}
\usepackage[percent]{overpic}
\usepackage[nolist,nohyperlinks]{acronym}
\usepackage{ulem}
\usepackage{xr}
\usepackage{hyperref}
\hypersetup{
	pdftoolbar=true,
	pdfmenubar=true,
	pdffitwindow=false,
	pdfstartview={FitH},
	pdftitle={},
	pdfauthor={},
	pdfsubject={},
	pdfcreator={},
	pdfproducer={},
	pdfnewwindow=true,
	colorlinks=true,
	linkcolor=black,
	citecolor=blue,
	filecolor=magenta,
	urlcolor=blue
}

\graphicspath{{figures/}}

\newcommand{\OO}[1]{\mathcal{O}\paren{#1}}
\newcommand{\paren}[1]{\left( #1 \right)}

\newcommand{\wh}[1]{\widehat{#1}}

\newcommand{\omax}{\omega_{\max}}
\newcommand{\sro}{Sr$_2$RuO$_4$}
\newcommand{\simp}{\Sigma_{\mathrm{imp}}}
\newcommand{\gimp}{G_{\mathrm{imp}}}
\newcommand{\gloc}{G_{\mathrm{loc}}}

\newcommand{\CCQ}{Center for Computational Quantum Physics, Flatiron Institute, 162 5th Avenue, New York, NY 10010, USA}
\newcommand{\CCM}{Center for Computational Mathematics, Flatiron Institute, 162 5th Avenue, New York, NY 10010, USA}
\newcommand{\UChicago}{Department of Chemistry, University of Chicago, 5735 S Ellis Ave, Chicago, IL 60637, USA}

\begin{document}

\title{Low rank Green's function representations applied to dynamical mean-field theory}

\author{Nan Sheng}
\email{nansheng@uchicago.edu}
\affiliation{\UChicago}
\affiliation{\CCQ}

\author{Alexander Hampel}
\affiliation{\CCQ}

\author{Sophie Beck}
\affiliation{\CCQ}

\author{Olivier~Parcollet}
\affiliation{\CCQ}
\affiliation{Universit\'e Paris-Saclay, CNRS, CEA, Institut de Physique Th\'eorique, 91191, Gif-sur-Yvette, France}

\author{Nils Wentzell}
\affiliation{\CCQ}

\author{Jason Kaye}
\email{jkaye@flatironinstitute.org}
\affiliation{\CCQ}
\affiliation{\CCM}

\author{Kun Chen}
\email{kunchen@flatironinstitute.org}
\affiliation{\CCQ}

\begin{abstract}
  Several recent works have introduced highly compact representations of single-particle 
  Green's functions in the imaginary time and Matsubara frequency
  domains, as well as efficient interpolation grids used to recover the
  representations. In particular, the intermediate representation with sparse
  sampling and the discrete Lehmann representation (DLR) make use of low rank
  compression techniques to obtain optimal approximations with controllable
  accuracy. We consider the use of the DLR in dynamical mean-field theory (DMFT)
  calculations, and in particular show that the standard full
  Matsubara frequency grid can be replaced by the compact grid of DLR Matsubara
  frequency nodes. We test the performance of the method for a DMFT calculation
  of Sr$_2$RuO$_4$ at temperature \SI{50}{K} using a continuous-time quantum Monte Carlo impurity
  solver, and demonstrate that Matsubara frequency quantities can be represented
  on a grid of only $36$ nodes with no reduction in accuracy, or increase in the
  number of self-consistent iterations, despite the presence of significant
  Monte Carlo noise.
\end{abstract}

\maketitle

\begin{acronym}
  \acro{QMC}{quantum Monte Carlo}
  \acro{CTQMC}{continuous-time quantum Monte Carlo}
  \acro{CTHYB}{continuous-time hybridization expansion}
  \acro{TMO}{transition metal oxides}
  \acro{DLR}{discrete Lehmann representation}
  \acro{IR}{intermediate representation}
  \acro{CSC}{charge self-consistent}
  \acro{BZ}{Brillouin zone}
  \acro{DFT}{density functional theory}
  \acro{DMFT}{dynamical mean-field theory}
  \acro{FT}{Fourier transform}
  \acro{KS}{Kohn-Sham}
  \acro{MIT}{metal-insulator transition}
  \acro{MLWF}{maximally localized Wannier function}
  \acro{OS}{one-shot}
  \acro{QE}{\textsc{Quantum~ESPRESSO}}
  \acro{TB}{tight-binding}
  \acro{W90}{\textsc{Wannier90}}
  \acro{WF}{Wannier function}
  \acro{AIM}{Anderson impurity model}
  \acro{PLO}{projected atomic orbitals}
\end{acronym}

\section{Introduction}

In the past several decades, \ac{DMFT}~\cite{Georges:1996} has become a
standard method for studying interacting fermionic lattice problems. In combination with first-principles
methods~\cite{Kotliar:2006,Held:2007}, it has
been widely adopted to calculate properties of strongly
correlated materials.
In such \ac{DMFT} calculations of real materials, the low
temperature regime is of particular importance, as numerous experimental examples show:
the critical temperature $T_\mathrm{C}$ for
superconductivity in \sro{} is as low as approximately \SI{1.5}{K} $\approx 10^{-4} \text{eV}$~\cite{Liu:2015}; the magnetic ordering in double-perovskite iridates sets is below
\SI{2}{K}~\cite{Terzic:2017}. In these cases, the ordering temperature energy scale
differs by about five orders of magnitude from the high energy cutoff of
approximately \SI{10}{eV}.

The single-particle Green's function, a central quantity in \ac{DMFT},
is often calculated in the imaginary time or Matsubara frequency
domain. The standard representation on an equispaced grid in imaginary
time, or on Matsubara frequencies up to a cutoff, is low-order accurate,
and requires
\begin{equation}
  N = \OO{\beta \omax}
\end{equation}
degrees of freedom. Here, $\beta$ is the inverse temperature, and $\omax$ is the
high energy cutoff of the spectral function (i.e., $\rho(\omega) = 0$ outside
$[-\omax,\omax]$). In typical \ac{DMFT} calculations, computing the local
Green's function requires a possibly expensive \ac{BZ} integration for each Matsubara frequency grid
point and each iteration of a self-consistency loop determining the chemical
potential. This cost can become substantial as the temperature is decreased.

A significant research effort has recently focused on developing compact and generic
representations of imaginary time and Matsubara frequency Green's functions,
beginning with orthogonal polynomial bases
\cite{boehnke11,kananenka16,gull18,dong20} and adaptive grid representations in
imaginary time \cite{ku02,kananenka16}. More recently, optimized basis sets
obtained from low rank compression of the Lehmann integral representation have
been developed, along with associated stable interpolation grids allowing
recovery of Green's functions from a small number of samples in either the
imaginary time or Matsubara frequency domains. This began with the introduction
of the orthogonal \ac{IR} basis~\cite{shinaoka17,chikano18}. Interpolation grids
for the \ac{IR} were later developed using the sparse sampling
method~\cite{li20}. Recently, some of the authors introduced the
\ac{DLR}~\cite{kaye22_dlr}, which uses a non-orthogonal but explicit basis of
exponentials, with associated \ac{DLR} interpolation grids. Both the \ac{IR} and
\ac{DLR} bases, and their interpolation grids, contain only 
\begin{equation}
  N = \OO{\log(\beta \omax) \log(\epsilon^{-1})}
\end{equation}
degrees of freedom, with $\epsilon$ a user-provided error tolerance. They
therefore yield exceptionally compact representations with controllable,
high-order accuracy. Fortran, Python, and Julia libraries are available for both
the \ac{IR} with sparse sampling \cite{wallerberger23} and the \ac{DLR} \cite{kaye22_libdlr}.
Low rank Green's function representations have been used to solve
self-consistent diagrammatic equations in a variety of applications, including
the SYK model~\cite{kaye22_dlr,kaye22_libdlr,kaye21_eqdyson},
the self-consistent finite temperature $GW$ method~\cite{li20, yeh22},
Eliashberg-type equations for superconductivity~\cite{wang20sc, cai22,
wang22, shinaoka22sc}, and Bethe-Salpeter-type equations for Hubbard
models~\cite{markus21bse}.

In this work we investigate the applicability and robustness of the \ac{DLR} in
self-consistent \ac{DMFT} calculations. Specifically, we replace the standard
Matsubara frequency grid with the compact \ac{DLR} grid in the
calculation of the local Green's function and all subsequent expressions in the
\ac{DMFT} equations.
We find that this method is stable, even in the presence of noisy
Green's function data as obtained from \ac{CTQMC} impurity solvers,
and that neither the convergence nor the
accuracy of self-consistent iteration is compromised. We demonstrate a reduction in computational effort and
memory required to calculate the local Green's function by over two orders of magnitude for the correlated Hund's metal
\sro{} at $T=\SI{50}{K}$.
Although the expensive solution of the impurity problem remains a barrier in many
\ac{DMFT} calculations, our approach therefore dramatically reduces the other
significant cost in the DMFT loop, and leads to a more automated procedure.

\section{Background}

\subsection{The dynamical mean-field theory loop}
\label{sec:dmft}

\begin{figure}
  \centering
  \includegraphics[width=1.0\linewidth]{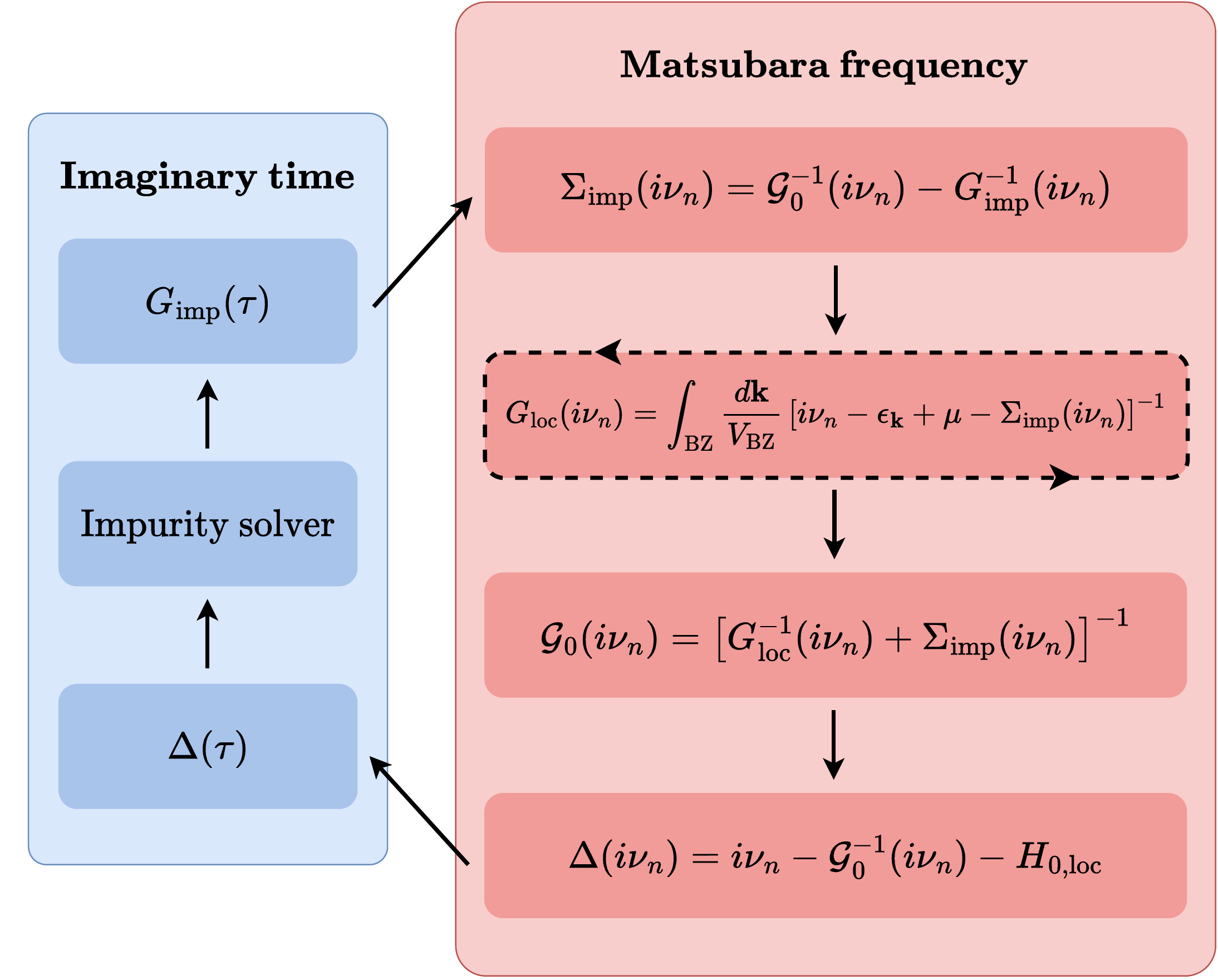}
   \caption{The steps of the DMFT loop. The arrows around the formula for
   $\gloc$ indicate that this quantity is computed self-consistently with the
   chemical potential to maintain the correct particle density. Our approach
   improves the efficiency of the DMFT loop by making two simple changes
   compared with the standard algorithm: (1) All
  operations in the Matsubara frequency domain are  carried out only at the \ac{DLR} nodes $\nu_n = \nu_{n_k}$, rather than the full Matsubara
  frequency grid, and (2) the imaginary time hybridization function
  $\Delta(\tau)$ is obtained from the computed values
  $\Delta(i \nu_{n_k})$ by forming a DLR expansion and obtaining its Fourier
  transform analytically.}
  \label{fig:dmftloop}
\end{figure}

We briefly outline the \ac{DMFT} equations, and refer the reader to
Refs.~\cite{Georges:1996,Kotliar:2006} for a more comprehensive overview.
The central quantity of interest is the local Green's function,
\begin{align}\label{eq:gloc}
   & \gloc(i\nu_n)= \int_{\mathrm{BZ}} \frac{d\mathbf{k}}{V_{\mathrm{BZ}}} \, {\left[i\nu_n-\epsilon_\mathbf{k}+\mu-\Sigma(\mathbf{k},i\nu_n)\right]}^{-1}.
\end{align}
Here $i \nu_n = i (2n+1) \pi / \beta$ is the Matsubara frequency variable (for
fermionic Green's functions), $\epsilon_\mathbf{k}$ is the non-interacting lattice Hamiltonian,
$\mu$ is the chemical potential, $\Sigma(\mathbf{k},i\nu_n)$ is the lattice
self-energy, and $V_{\mathrm{BZ}}$ is the volume of the \ac{BZ}. The chemical
potential can be computed self-consistently in each \ac{DMFT} iteration in order to
maintain the correct particle density.
In \ac{DMFT}, the self-energy is approximated as a local quantity,
and is computed from the Green's function $\gimp(i\nu_n)$ of an effective
impurity problem via the Dyson equation
\begin{align}\label{eq:sigma}
   & \Sigma(\mathbf{k},i\nu_n) \approx \simp(i\nu_n) = \mathcal{G}_0^{-1}(i\nu_n) - G_{\mathrm{imp}}^{-1}(i\nu_n)\,.
\end{align}
Here, the effective non-interacting bath is represented by the Weiss mean-field,
\begin{align}\label{eq:gnod}
    & \mathcal{G}_0^{-1}(i\nu_n) = G_{\mathrm{loc}}^{-1}(i\nu_n) + \simp(i\nu_n)\,,
\end{align}
obtained from the local Green's function \eqref{eq:gloc}.
The local Green's function
is obtained self-consistently, and convergence is reached when
$\gloc = \gimp$.
For concreteness, we focus in this paper on the \ac{CTHYB} impurity solver~\cite{Seth2016274}, in which the
impurity problem is defined through the local
non-interacting Hamiltonian
\begin{align}
   H_{0,\mathrm{loc}} = \int_{\mathrm{BZ}} \frac{d\mathbf{k}}{V_{\mathrm{BZ}}}
   \, \epsilon_\mathbf{k} - \mu
\end{align}
and the Matsubara frequency hybridization function
\begin{align}\label{eq:hyb}
   \Delta(i\nu_n) = i\nu_n - \mathcal{G}_0^{-1}(i\nu_n) - H_{0,\mathrm{loc}},
\end{align} or its Fourier transform $\Delta(\tau)$ to the imaginary time domain.
We emphasize, however, that compact representations of the type used here are in
principle equally applicable for other types of impurity solvers.

The \ac{DMFT} loop, outlined above, is summarized in Fig. \ref{fig:dmftloop}.
Although the solution of
the impurity problem is often the most computationally intensive and technical
step in the \ac{DMFT} loop, it is outside the scope of our current discussion.
Rather, we focus on the calculation of $\gloc(i
\nu_n)$, which requires the evaluation of a \ac{BZ} integral for each Matsubara
frequency grid point $i\nu_n$. In typical calculations all Matsubara frequency
points are used up to a cutoff $\OO{\omax}$ (yielding $\OO{\beta \omax}$ points
in total), in order
to capture the effective energy scales of the system. We demonstrate here that the number of
Matsubara frequency points at which $\gloc(i \nu_n)$ must be computed can be
dramatically reduced.

\subsection{Discrete Lehmann representation and compact Matsubara frequency
grids}

The \ac{DLR} method provides a compact, explicit basis for Matsubara
Green's functions and self-energies, along with associated interpolation grids.
We give a brief review of these
concepts here, and refer to
Ref.~\onlinecite{kaye22_dlr} for a detailed presentation and analysis.

Each Matsubara Green's function $G(i \nu_n)$ has a spectral Lehmann representation
\begin{equation} \label{eq:lehmann}
G(i \nu_n) = \int_{-\infty}^\infty K(i \nu_n,\omega) \, \rho(\omega) \, d\omega,
\end{equation}
where $\rho(\omega)$ is the spectral function, and the analytic continuation
kernel $K$ is given by
\begin{equation} \label{eq:kmf}
  K(i\nu_n,\omega) \equiv (i\nu_n-\omega)^{-1}.
\end{equation}
In most practical applications, $\rho$ is unknown, but $G(i \nu_n)$ can either be
sampled directly or obtained from samples of the imaginary time Green's
function $G(\tau)$. We assume $\rho$ can be truncated beyond a frequency cutoff
$\abs{\omega} = \omax$. Defining the dimensionless parameter
\[\Lambda \equiv \beta \omax,\]
and nondimensionalizing variables by $\nu_n \gets \beta \nu_n$ and $\omega \gets
\beta \omega$, we obtain
the truncated
Lehmann representation
\begin{equation} \label{eq:lehmanntrunc}
  G(i \nu_n) = \int_{-\Lambda}^\Lambda K(i \nu_n,\omega) \, \rho(\omega) \, d\omega,
\end{equation}
where $\nu_n$ is given as above with $\beta = 1$, and the arguments of $G$, $\rho$ have been
suitably rescaled. 

It can be
shown that the kernel of this integral representation, $K(i \nu_n,\omega)$,
has super-exponentially decaying singular values
\cite{shinaoka17,chikano18}. This low rank structure is indicative of the
well-known ill-conditioning of analytic continuation from the Matsubara Green's function
to the spectral function on the real
frequency axis~\cite{Silver/Sivia/Gubernatis:1990}. However, it is advantageous for the
representation of Matsubara Green's functions themselves, implying
that $K(i \nu_n,\omega)$ can be approximated
for any $\omega \in [-\Lambda,\Lambda]$ as a linear combination of a
small number of basis functions. In particular, the DLR approach uses frequency samples
of the kernel itself as basis functions:
\begin{equation} \label{eq:kapprox}
  K(i \nu_n,\omega) \approx \sum_{l=1}^r K(i \nu_n,\omega_l) \pi_l(\omega).
\end{equation}
The $r$ \textit{DLR frequencies} $\omega_l$ can be selected
automatically by the pivoted Gram-Schmidt algorithm such that the approximation
in \eqref{eq:kapprox} is numerically stable, and accurate to a
user-provided error tolerance \cite{cheng05}. Substitution of \eqref{eq:kapprox} into
\eqref{eq:lehmanntrunc} demonstrates the existence of an expansion of an
arbitrary Matsubara Green's function in the basis
$K(i \nu_n,\omega_l)$,
\begin{equation} \label{eq:gdlrmf}
  G(i \nu_n) \approx \sum_{l=1}^r K(i \nu_n,\omega_l) \wh{g_l},
\end{equation}
with $\wh{g_l} = -\int_{-\Lambda}^{\Lambda} \pi_l(\omega) \rho(\omega) \,
d\omega$.

The rapid decay of the singular values of $K$ implies the scaling $r = \OO{\log
\paren{\Lambda} \log\paren{\varepsilon^{-1}}}$, yielding exceptionally compact expansions
at high accuracies and low temperatures. For example, Matsubara Green's functions
with $\Lambda = 100$ can be represented to 6-digit accuracy by fewer
than $20$ basis functions; with $\Lambda = 10^4$ to 6-digit accuracy by fewer than $50$
basis functions; and with $\Lambda = 10^6$ to 10-digit accuracy by fewer
than $120$ basis functions. By contrast, in a typical calculation, for example with $\beta =
\SI{1000}{eV}^{-1}$ and $\omax = \SI{10}{eV}$ ($\Lambda = 10^4$), one would typically require on the order of tens of
thousands of Matsubara frequencies.
We emphasize that given $\Lambda$ and
$\varepsilon$, the representation is universal; that is,
independent of the specific structure of the spectral function $\rho$
characterizing the Green's function, which is already taken into account
by the automatic compression of the kernel $K$.

Since $\rho$ is typically not known and the \textit{DLR
coefficients} $\wh{g_l}$ cannot be
computed directly, they can in practice be recovered by fitting, or by
interpolation at a collection of $r$
\textit{DLR Matsubara frequency nodes} $\{i \nu_{n_k}\}_{k=1}^r$
\cite{kaye22_dlr}. These nodes can
be obtained automatically, using a process similar to that used to
obtain the DLR frequencies, to ensure stable interpolation. Thus, a Green's
function $G$ can be characterized, to within a controllable error, by its values
$G(i \nu_{n_k})$ at the DLR nodes.

The Fourier transform of \eqref{eq:gdlrmf} yields an imaginary time representation,
\begin{equation} \label{eq:gdlrit}
  G(\tau) \approx \sum_{l=1}^r K(\tau,\omega_l) \wh{g_l},
\end{equation}
with
\[K(\tau,\omega) \equiv \frac{e^{-\omega \tau}}{1+e^{-\omega}}\]
in the transformed variables $\tau \gets \tau/\beta$, $\omega \gets \beta
\omega$.
As for the Matsubara frequency expansion, $G(\tau)$ can either be recovered
by least squares fitting, or by interpolation at a collection of
automatically selected \textit{DLR imaginary time nodes} $\{\tau_k\}_{k=1}^r$.
We note that the DLR interpolation procedure is similar to the method of sparse
sampling used in conjunction with the IR basis, in which interpolation nodes are
selected based on the extrema of the highest degree IR basis function \cite{li20}.

\section{Restriction to compact Matsubara frequency grid}

We propose the following procedure to improve the efficiency of the \ac{DMFT} loop: Given
the self-energy $\simp$, the local Green's function $\gloc$ is
evaluated only at the $r$ DLR Matsubara frequency nodes $\{i \nu_{n_k}\}_{k=1}^r$, as are the Weiss
mean-field $\mathcal{G}_0$ from \eqref{eq:gnod} and the hybridization
function $\Delta$ from \eqref{eq:hyb}. At this point, the DLR expansion
of $\Delta(i \nu_n)$ is formed by interpolation from its values at the DLR nodes using the
representation in \eqref{eq:gdlrmf}, with $G$ replaced by $\Delta$. $\Delta(\tau)$ is then given analytically by a DLR expansion
in imaginary time, as in \eqref{eq:gdlrit}. The
rest of the DMFT procedure can be carried out without modification.

\begin{figure*}[t]
  \centering
  \includegraphics[width=1.0\linewidth]{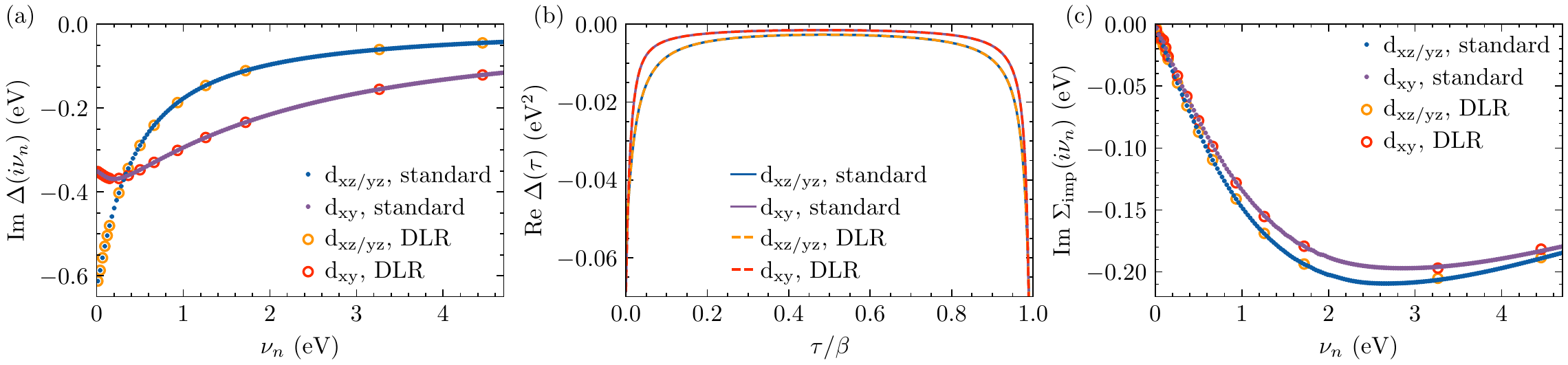}
  \caption{Hybridization function $\Delta$ and self-energy $\simp$ during the first iteration of the
  DMFT loop for the \sro{} example, demonstrating the use of the \ac{DLR} procedure.
  (a) $\Delta(i \nu_n)$ (from initial guess with zero self-energy) given by
  \eqref{eq:hyb}, with DLR nodes indicated.
  (b) $\Delta(\tau)$ obtained using the standard method, i.e. asymptotic
  expansion and discrete Fourier transform, and DLR interpolation from the values $\Delta(i \nu_{n_k})$.
  (c)
  $\simp(i
  \nu_n)$ calculated after the
  impurity problem is solved with the hybridization function obtained using both
  methods.}
  \label{fig:SRO_DLR_DMFT_loop}
\end{figure*}

The primary purpose of this paper is to verify that systematic or statistical error generated by the
quantum impurity solver does not destabilize our proposed procedure. Although
this question depends on the specific choice of impurity solver, we carry out
tests using the most popular solver, \ac{CTQMC}. We show in the next section
that the interpolation procedure is stable to Monte Carlo noise, and that the
convergence of the DMFT loop is not affected by the reduction of the Matsubara
frequency grid.

\section{Numerical example: Strontium Ruthenate}

\begin{figure}
  \centering
  \includegraphics[width=1.0\linewidth]{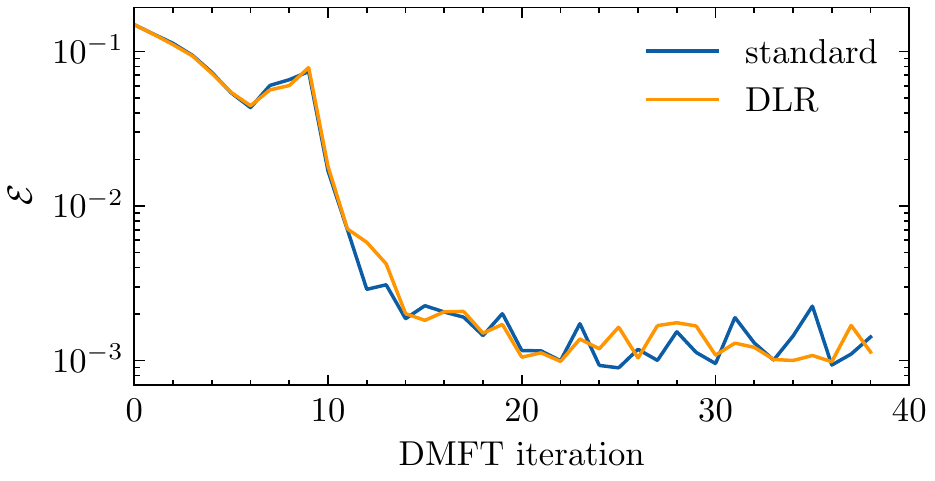}
  \caption{Convergence of DMFT self-consistency for the \sro{} example using
  standard and DLR procedures, measured using \eqref{eq:error}.}
  \label{fig:SRO_l2_conv}
\end{figure}

\begin{figure*}
  \centering
  \includegraphics[width=0.9\linewidth]{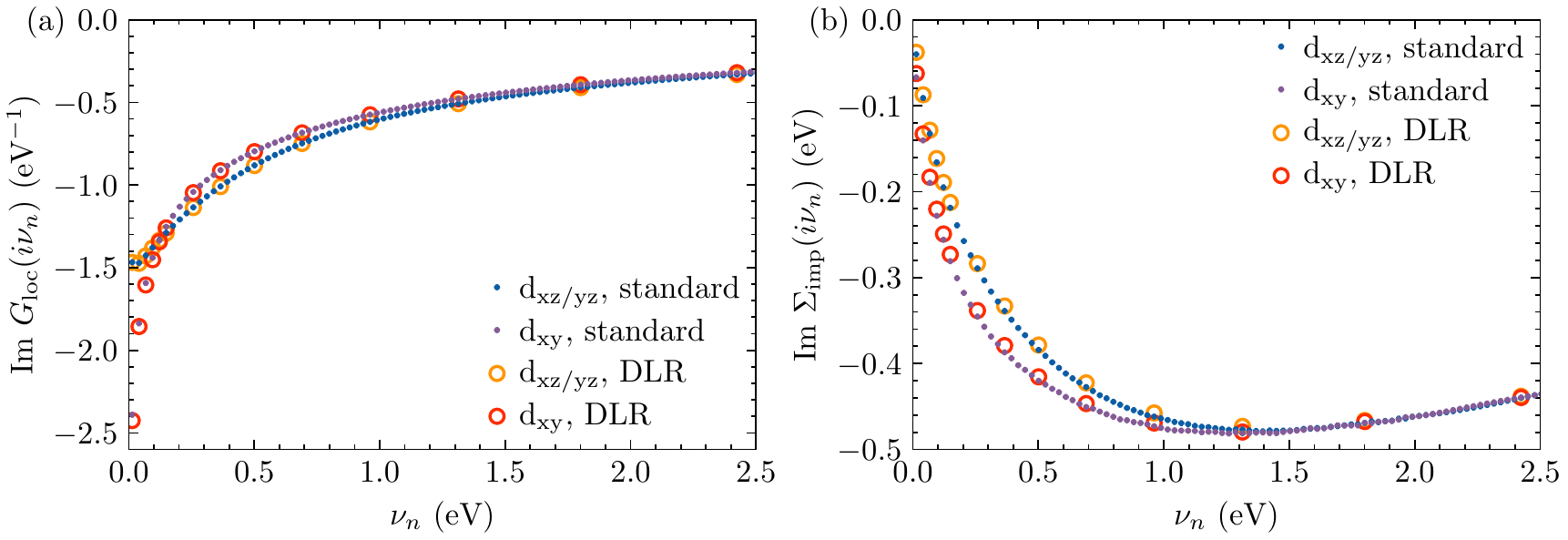}
  \caption{Converged results for the \sro{} example using standard
  and DLR procedures: (a) local Green's function, and (b) impurity
  self-energy in a low-frequency window.}
  \label{fig:SRO_final}
\end{figure*}

We demonstrate our procedure using the
correlated Hund's metal Sr$_2$RuO$_4$~\cite{Georges/Medici/Mravlje:2013} at low $T$. We compute the electronic
structure using the planewave-based \textsc{Quantum~ESPRESSO}
package~\cite{Giannozzi_et_al:2009} using the standard Perdew–Burke–Ernzerhof
exchange-correlation functional, and scalar-relativistic ultrasoft
pseudopotentials~\cite{Garrity_et_al:2014}. After structural optimization on a $12\times12\times12$ Monkhorst-Pack grid, we obtain lattice
parameters that correspond to $a=\SI{3.880}{\mbox{\AA}}$
and $c=\SI{12.887}{\mbox{\AA}}$ in the conventional unit cell (space group I4/mmm
(139)). The primitive unit cell
contains one ruthenium site with a partially filled $t_{2g}$ shell for which
we construct three maximally localized Wannier functions~\cite{wannier90},
representing the degenerate $d_{xz}$/$d_{yz}$ orbitals and the $d_{xy}$ orbital.
We recompute the Hamiltonian on a $40\times40\times40$ $k$-point grid using
Wannier interpolation in order to compute the \ac{BZ} integrals in
\eqref{eq:gloc}
by equispaced integration. We add a local rotationally invariant Hubbard-Kanamori
interaction with $U=\SI{2.3}{eV}$ and $J=\SI{0.4}{eV}$~\cite{Georges/Medici/Mravlje:2013}.
The impurity problem is solved using TRIQS/CTHYB~\cite{Seth2016274} in the TRIQS
library~\cite{parcollet15}. To address
the well-known numerical instability of computing the self-energy via the Dyson
equation in \eqref{eq:sigma} in the presence of \ac{QMC}
noise, we replace this formula at high frequencies with an asymptotic expansion.
This expansion is given by a polynomial in $(i \nu_n)^{-1}$, fit to
$\simp$ in a window in which the \ac{QMC}
noise is sufficiently small so that \eqref{eq:sigma} is valid.

The \ac{DMFT} calculation is implemented using the TRIQS
library~\cite{parcollet15}, and the Python library \texttt{pydlr} provided by
\texttt{libdlr}~\cite{kaye22_libdlr,libdlr} is used for DLR calculations. We
solve the \ac{DMFT} equations at $\beta=\SI{232}{eV}^{-1}$, which corresponds to
$T=\SI{50}{K}$. At this temperature, without the \ac{DLR}, approximately 12000
Matsubara frequency nodes are required to adequately capture the slowly-decaying
tail of the Green's functions to allow for accurate Fourier transforms. 
More specifically, in the TRIQS library, the Fourier transform
$\Delta(\tau)$ of $\Delta(i \nu_n)$ is obtained by the following procedure: (1)
fit an asymptotic expansion in inverse powers of $i \nu_n$ to $\Delta(i \nu_n)$,
(2) Fourier transform this asymptotic expansion analytically, (3) Fourier
transform the difference between $\Delta(i \nu_n)$ and its asymptotic expansion,
which is rapidly decaying, by a discrete Fourier transform on the Matsubara
frequency grid, and add the results. 
Choosing $\omax = \SI{12}{eV}$ and $\epsilon = \SI{e-6}{eV}^{-1}$, the number of DLR basis functions and Matsubara frequency
nodes is $r = 36$, reducing the number of \ac{BZ} integrals required to
calculate $\gloc$ in \eqref{eq:gloc} by a factor of over $300$. Furthermore, our approach
avoids the complicated Fourier transform procedure used with the standard
Matsubara frequency grid, since 
$\Delta(\tau)$ is obtained from the DLR expansion of $\Delta(i \nu_n)$ by
analytical Fourier transform.

Fig. \ref{fig:SRO_DLR_DMFT_loop} shows the first iteration of the DMFT loop
comparing the hybridization function and the self-energy obtained using the
standard method and the DLR approach (shown in markers/dashed lines
and solid lines, respectively).
In our scheme, we first compute $\gloc$,
$\mathcal{G}_0$, and $\Delta$ at the DLR nodes $i\nu_n = i\nu_{n_k}$,
using zero self-energy as an initial guess in \eqref{eq:gloc}.
The hybridization function $\Delta(i \nu_n)$ is shown in Fig. \ref{fig:SRO_DLR_DMFT_loop}a on the full
Matsubara frequency grid, as used in the standard method, with the DLR nodes
used in our method indicated. The DLR expansion of
$\Delta(\tau)$ obtained from interpolation at these nodes and analytical Fourier
transform is shown in Fig. \ref{fig:SRO_DLR_DMFT_loop}b. Since no Monte Carlo noise has
been introduced at this stage, the DLR expansion of $\Delta(\tau)$ is correct to the DLR tolerance
$\epsilon$. We then solve the impurity problem using the DLR
expansion of $\Delta(\tau)$ to obtain the impurity Green's function, and
subsequently the self-energy $\Sigma(i \nu_n)$, shown in Fig.
\ref{fig:SRO_DLR_DMFT_loop}c. We see that the self-energies obtained using the
hybridization function obtained using the full grid \ac{DMFT} procedure (shown at all Matsubara frequencies as dots)
and the DLR procedure (shown at the DLR nodes as open circles) in the impurity
solver agree to within the Monte Carlo noise level. 

We next run the standard and modified DMFT loops until self-consistency.
Convergence is measured by monitoring the quantity
\begin{equation} \label{eq:error}
\mathcal{E} = \sqrt{\frac{1}{\beta} \int_0^\beta d\tau \,
\norm{\gimp(\tau) - \gloc(\tau)}_F^2},
\end{equation}
where $\norm{\cdot}_F$ indicates the
Frobenius norm, and the normalization prevents a trivial scaling of the
error with $\beta$, assuming a uniform distribution of Monte Carlo error.
Fig. \ref{fig:SRO_l2_conv} shows that the convergence behavior is nearly
identical for the two approaches, with both reaching self-consistency after
after approximately 20 iterations. Finally, Fig.~\ref{fig:SRO_final}
shows $\gloc(i \nu_n)$ and $\simp(i\nu_n)$ at convergence, demonstrating that
the final results of the two calculations agree to within the Monte Carlo noise
level.

\section{Conclusion}\label{sec:Conclusion}

Our proposed method improves the efficiency of the \ac{DMFT} procedure by
replacing the
standard full Matsubara frequency grid with a highly compact grid compatible
with interpolation using the DLR basis. We demonstrate the effectiveness of this
approach for a \ac{DMFT} calculation of \sro{} using \ac{CTQMC} as the impurity
solver. In general, our results suggest that the standard representations of
quantities appearing in the \ac{DMFT} loop can be replaced by much more
efficient representations, such as the \ac{DLR}, without incurring a penalty in
accuracy or stability.
We note that the same approach should be applicable to other impurity solvers,
in particular fast approximate solvers used in real materials applications.

\acknowledgments 
The Flatiron Institute is a division of the Simons Foundation.

\bibliographystyle{ieeetr}
\bibliography{dlrdmft}

\end{document}